\documentclass[prd,preprintnumbers,amsmath,amssymb]{revtex4}

\input epsf
\begin{document}
\draft

\title{Charged particles' tunnelling from the Kerr-Newman black hole}
\author{Jingyi Zhang\footnote{E-mail: physicz@tom.com} and Zheng Zhao\footnote{{E-mail:
zhaoz43@hotmail.com}}} \affiliation{ Department of Physics ,
Beijing Normal University, Beijing 100875, China}
\date{\today}

\begin{abstract}
In this letter, Parikh-Wilczek tunnelling framework, which treats
Hawking radiation as a tunnelling process, is extended, and the
emission rate of a charged particle tunnelling from the
Kerr-Newman black hole is calculated. The emission spectrum takes
the same functional form as that of uncharged particles and
consists with an underlying unitary theory but deviates from the
pure thermal spectrum. Moreover, our calculation indicates that
the emission process is treated as a reversible process in the
Parikh-Wilczek tunnelling framework, and the information
conservation is a natural result of the first law of black hole
thermodynamics.\\\\PACS number(s): 04.70.Dy
\end{abstract}
\maketitle
\vskip2pc
\section{Introduction}
In 2000, Parikh and Wilczek proposed an approach to calculate the
emission rate at which particles tunnel across the event
horizon\cite{Parikh1}. They treat Hawking radiation as a
tunnelling process, and the WKB method is
used\cite{Parikh2,Parikh3}. In this way a corrected spectrum,
which is accurate to a first approximation, is given. Their
results are considered to be in agreement with an underlying
unitary theory. Following this method, a lot of static or
stationary rotating black holes are studied
\cite{Hemming,Medved,Alves,Vagenas1,Vagenas2,Vagenas3,Vagenas4,Zhang1,Zhang2,Zhang3,Zhang4}.
The same results, that is, Hawking radiation is no longer pure
thermal, unitary theory is satisfied and information is conserved,
are obtained. In this letter, we extend the Parikh-Wilczek
tunnelling framework to calculate the emission rate of a charged
particle from the Kerr-Newman black hole. We will treat the
charged particle as a de Broglie wave. There are two difficulties
to overcome. The first is that in order to do a computation, one
need to find the equation of motion of a charged massive
tunnelling particle. The second is how to take into account the
effect of the electromagnetic field outside the hole. In fact, the
first problem is solved in Ref. \cite{Zhang3}. In similar manner
we can obtain the concrete expression of the equation of motion.
About the second problem, we find that in our discussion the
Lagrangian density of the electromagnetic field can be expressed
by a set of generalized coordinates $A_\mu=(A_t,\,A_1\,A_2\,A_3)$.
But these coordinates are ignorable coordinates in dragged
coordinate system. To eliminate the freedoms corresponding to
these coordinates, we modify the Lagrangian function and use the
WKB method to get the corrected emission spectrum.

The remainder of the letter is organized as follows. In section 2,
we introduce the Painlev$\acute{\mathrm{e}}$-Kerr-Newman
coordinate system, calculate the phase velocity of the de Broglie
wave, and therefore obtain the equation of motion of a charged
particle. In section 3, we modify the Lagrangian function and use
Parikh-Wilczek tunnelling framework to get the corrected emission
spectrum. Finally, in section 4, we present a short discussion
about our result. Throughout the paper, the geometrized units
$(G\equiv c\equiv\hbar\equiv 1)$ are used.
\section{Phase velocity and electromagnetic potential}
The line element of the Kerr-Newman black hole can be written in
the
form\cite{Newman}\cite{Wald}%
\begin{align}
ds^{2}=-(  &
1-\frac{2Mr-Q^{2}}{\rho^{2}})dt_{k}^{2}+\frac{\rho^{2}}{\Delta
}dr^{2}+\rho^{2}d\theta^{2}+[(r^{2}+a^{2})\sin^{2}\theta+\frac{(2Mr-Q^{2}%
)a^{2}\sin^{4}\theta}{\rho^{2}}]d\varphi^{2}\nonumber\\
&
-\frac{2(2Mr-Q^{2})a\sin^{2}\theta}{\rho^{2}}dt_{k}d\varphi,\label{Kerr-Newman
element}
\end{align}

\bigskip where
\begin{align}
\rho^2 &  \equiv r^{2}+a^{2}\cos^{2}\theta,\\
\Delta &  \equiv r^{2}+a^{2}+Q^{2}-2Mr.
\end{align}
The event horizon $r=r_{H}$ is given by%
\begin{equation}
r_{H}=M+\sqrt{M^{2}-a^{2}-Q^{2}},
\end{equation}
and the 4-dimensional electromagnetic potential is
\begin{equation}
A_a=-\rho^{-2}Qr[(dt)_a-a\sin^2\theta (d\varphi)_a].
\end{equation}
To calculate the emission rate, we should adopt the dragged
coordinate system\cite{Zhang2}. The line element in this
coordinate system is
\begin{align}
ds^{2}  &  =-\frac{\rho^{2}\Delta}{(r^{2}+a^{2})^{2}-\Delta
a^{2}\sin
^{2}\theta}dt_{d}^{2}+\frac{\rho^{2}}{\Delta}dr^{2}+\rho^{2}d\theta
^{2}\nonumber\\
&
=\widehat{g}_{00}dt_{d}^{2}+g_{11}dr^{2}+g_{22}d\theta^{2}.\label{drag}
\end{align}
where%
\begin{equation}
\widehat{g}_{00}=-\frac{\rho^{2}\Delta}{(r^{2}+a^{2})^{2}-\Delta
a^{2}\sin ^{2}\theta},
\end{equation}%
\begin{equation}
g_{11}=\frac{\rho^{2}}{\Delta},
\end{equation}%
\begin{equation}
g_{22}=\rho^{2}.
\end{equation}
Since
\begin{equation}
(\frac{\partial}{\partial t_d})^a=(\frac{\partial}{\partial
t_k})^a+\Omega (\frac{\partial}{\partial \varphi})^a,
\end{equation}
we can easily obtain the components of the electromagnetic
potential in the dragged coordinate system
\begin{equation}
A'_0=A_a(\frac{\partial}{\partial
t_d})^a=-\rho^{-2}Qr[1-a\Omega\sin^2\theta],\quad\quad
A'_1=A'_2=0,
\end{equation}
where $\Omega=-g_{03}/g_{33}$ is the dragged angular velocity. As
discussed in \cite{Zhang2}, there is a coordinate singularity in
the metric (\ref{drag}) at the radius of the event horizon. To
calculate the emission rate, we should introduce the
Painlev$\acute{\mathrm{e}}$-Kerr-Newmang coordinate system.

The line element in the Painlev$\acute{\mathrm{e}}$-Kerr-Newman
coordinate system is given in Ref. \cite{Zhang2}. Namely,
\begin{align}
ds^{2}  &  =\widehat{g}_{00}dt^{2}+2\sqrt{\widehat{g}_{00}(1-g_{11}%
)}dtdr+dr^{2}+[\widehat{g}_{00}G(r,\theta)^{2}+g_{22}]d\theta^{2}\nonumber
\\
&  +2\widehat{g}_{00}G(r,\theta)dtd\theta+2\sqrt{\widehat{g}_{00}(1-g_{11}%
)}G(r,\theta)drd\theta,\label{Painleve Kerr-Newman}
\end{align}
which is obtained from (\ref{drag}) by the coordinate
transformation
\begin{equation}
dt_{k}=dt+F(r,\theta)dr+G(r,\theta)d\theta, \label{dt}%
\end{equation}
where $F(r,\theta)$ satisfies
\begin{equation}
g_{11}+\widehat{g}_{00}F(r,\theta)^{2}=1, \label{demend}%
\end{equation}
and $G(r,\theta)$ is decided by
\begin{equation}
G(r,\theta)=\int\frac{\partial
F(r,\theta)}{\partial\theta}dr+C(\theta),
\end{equation}
here, $C(\theta)$ is an arbitrary analytic function of $\theta$.

The components of the electromagnetic potential in the
Painlev$\acute{\mathrm{e}}$-Kerr-Newman coordinate system is
\begin{equation}
A_0=-\rho^{-2}Qr[1-a\Omega\sin^2\theta],\quad\quad
A_1=A_2=0\label{a1}.
\end{equation}
From (\ref{a1}) we obtain the electromagnetic potential on the
event horizon
\begin{equation}
A_0|_{r_H}=-V_0=-\frac{Qr\!_H}{r^2_H+a^2},\quad\quad A_1=A_2=0.
\end{equation}
Similar to Ref. \cite{Zhang4}, we treat the charged particle as a
de Broglie wave and we can easily obtain the expression of
$\overset{.}{r}$. Namely,
\begin{equation}
\overset{.}{r}=v_{p}=-\frac{1}{2}\frac{\widehat{g}_{00}}{\widehat{g}_{01}}
=\frac{\Delta}{2}\sqrt{\frac{\rho^2}{(\rho^2-\Delta)[(r^{2}+a^{2})^{2}-\Delta
a^{2}\sin ^{2}\theta ]}}. \label{vp}
\end{equation}
Note that to calculate the emission rate correctly, we should take
into account the self-gravitation of the tunnelling particle with
energy $\omega$ and electric charge $q$. That is, we should
replace $M$ and $Q$ with $M-\omega$ and $Q-q$ in (\ref{Painleve
Kerr-Newman}) and (\ref{vp}), respectively.

\section{emission rate}
When we investigate a charged particle's tunnelling, the effect of
the electromagnetic field should be taken into account. That is,
the matter-gravity system consists of the black hole and the
electromagnetic field outside the hole. We write the Lagrangian
function of the matter-gravity
system as%
\begin{equation}
L=L_{m}+L_{e},
\end{equation}
where $L_{e}=-\frac{1}{4}F_{\mu\nu}F^{\mu\nu}$ is the Lagrangian
function of the electromagnetic field corresponding to the
generalized coordinates $A_{\mu}=(A_{t},\quad0,\quad0)$ in the
Painlev$\acute{\mathrm{e}}$-Kerr-Newman coordinate
system\cite{Makela}. When a charged particle tunnels out, the
system transit from one state to another. But from the expression
of $L_{e}$ we find that $A_{\mu}=(A_{t},\quad0,\quad0)$ is an
ignorable coordinate. Moreover, in the dragged coordinate system,
the coordinate $\varphi$ does not appear in the line element
expressions (\ref{drag}) and (\ref{Painleve Kerr-Newman}). That is
to say, $\varphi$ is also an ignorable coordinate in the
Lagrangian function $L$. To eliminate these two freedoms
completely, the action for the classically forbidden trajectory
should be written as
\begin{equation}
S=\int\nolimits_{t_{i}}^{t_{f}}(L-P_{A_{t}}\overset{.}{A_{t}}-P_{\varphi}\overset{.}{\varphi})dt,
\end{equation}
which is related to the emission rate of the tunnelling particle
by
\begin{equation}
\Gamma\sim e^{-2\operatorname{Im}S}.
\end{equation}
Therefore, the imaginary part of the action is
\begin{align}
\operatorname{Im}S&=\operatorname{Im}\{\int\nolimits_{r_{i}}^{r_{f}}
[P_{r}-\frac{P_{A_{t}}\overset{.}{A_{t}}}{\overset{.}{r}}-\frac{P_{\varphi}\overset{.}{\varphi}}{\overset{.}{r}}
]dr\}\nonumber\\&=\operatorname{Im}\{\int\nolimits_{r_{i}}^{r_{f}}[\int_{(0,0,0)}
^{(P_{r},P_{A_{t}},P_{\varphi})}dP_{r}^{^{\prime}}-\frac{\overset{.}{A_{t}}}{\overset
{.}{r}}dP_{A_{t}}^{^{\prime}}-\frac{\overset{.}{\varphi}}{\overset
{.}{r}}dP_{\varphi}^{^{\prime}}]dr\}, \label{ims1}
\end{align}
where $P_{A_{t}}$  and $P_{\varphi}$ are the canonical momentums
conjugate to $A_{t}$ and $\varphi$, respectively. If we treat the
black hole as a rotating sphere and consider the particle
self-gravitation, we have
\begin{equation}
\overset{.}{\varphi}=\Omega'_H,
\end{equation}
and
\begin{equation}
J'=(M-\omega ')a=P'_{\varphi},
\end{equation}
where $\Omega'_H$ is the dragged angular velocity of the event
horizon. The imaginary part of the action can be rewritten as
\begin{equation}
\operatorname{Im}S=\operatorname{Im}\{\int\nolimits_{r_{i}}^{r_{f}}[\int_{(0,0,J)}
^{(P_{r},P_{A_{t}},J-\omega
a)}dP_{r}^{^{\prime}}-\frac{\overset{.}{A_{t}}}{\overset
{.}{r}}dP_{A_{t}}^{^{\prime}}-\frac{\Omega'_H}{\overset
{.}{r}}dJ']dr\}. \label{ims2}
\end{equation}
We now eliminate the momentum in favor of energy by using
Hamilton's equations
\begin{equation}
\overset{.}{r}=\frac{dH}{dP_{r}}\mid_{(r;A_{t},P_{A_{t}};\varphi,P_{\varphi})}=\frac{d(M-\omega')}{dP_{r}}=\frac{dM'}{dP_{r}}, \label{r2}%
\end{equation}%
\begin{equation}
\overset{.}{A_{t}}=\frac{dH}{dP_{A_{t}}}\mid_{(A_{t};r,P_{r};\varphi,P_{\varphi})}=\frac{V'_0dQ'}{dP_{A_{t}}}=\frac{(Q-q')r_H}{r^2_H+a^2}\cdot\frac{d(Q-q')}{dP_{A_{t}}}. \label{r3}%
\end{equation}
Note that to derive (\ref{r3}) we have treated the black hole as a
charged conducting sphere\cite{Damour}.

Based on similar discussion to \cite{Parikh1,Parikh2}, it follows
directly that a charged particle tunnelling across the event
horizon sees the effective metric of Eq. (\ref{Painleve
Kerr-Newman}), although with the replacements $M\rightarrow
M-\omega'$ and $Q\rightarrow Q-q'$. The same substitutions in Eq.
(\ref{vp}) yield the desired expression of $\overset {.}{r}$ as a
function of $\omega'$ and $q'$. Thus, we can rewrite (\ref{ims2})
in the following explicit manner
\begin{equation}
\operatorname{Im}S=\operatorname{Im}\int\nolimits_{r_{i}}^{r_{f}}[\int
\frac{2\sqrt{(\rho^2-\Delta')[(r^{2}+a^{2})^{2}-\Delta' a^{2}\sin
^{2}\theta ]}}{\Delta'
\sqrt{\rho^2}}(dM'-\frac{Q'r'_H}{r'^2_H+a^2}dQ'-\Omega'_HdJ')]dr.
\label{ims3}
\end{equation}
where
\begin{equation}
\Delta^{^{\prime}}=r^{2}+a^{2}+Q'^{2}-2M'r=(r-r_{+}%
^{^{\prime}})(r-r_{-}^{^{\prime}}),
\end{equation}%
\begin{equation}
r'_{\pm}=(M-\omega')\pm\sqrt{(M-\omega')^{2}-a^{2}-(Q-q')^{2}},
\end{equation}
\begin{equation}
r_{i}=M+\sqrt{M^{2}-a^{2}-Q^{2}},
\end{equation}%
\begin{equation}
r_{f}=M-\omega+\sqrt{(M-\omega)^{2}-a^{2}-(Q-q)^{2}}.
\end{equation}
We see that
$r=r'_{+}=(M-\omega')+\sqrt{(M-\omega')^{2}-a^{2}-(Q-q')^{2}}$ is
a pole. The integral can be evaluated by deforming the contour
around the pole, so as to ensure that positive energy solution
decay in time. Doing the $r$ integral first we obtain
\begin{equation}
\operatorname{Im}S=-\frac{1}{2}\int\nolimits_{(M,Q)}^{(M-\omega,Q-q)}
\frac{4\pi(M'^2+M'\sqrt{M'^{2}-a^{2}-Q'^{2}}-\frac{1}{2}Q'^2)}{\sqrt{M'^{2}-a^{2}-Q'^{2}}}(dM'-\frac{Q'r'_H}{r'^2_H+a^2}dQ'-\Omega'_HdJ')dr.
\label{ims4}
\end{equation}
Finishing the integration we get
\begin{eqnarray}
\operatorname{Im}S\!&=&\!\pi[M^2-(M-\omega)^2+M\sqrt{M^2-a^2-Q^2}-(M-\omega)\sqrt{(M-\omega)^2-a^2-(Q-q)^2}-\frac{1}{2}(Q^2-(Q-q)^2)]\\&=&\!-\frac{1}{2}\Delta
S_{BH}.\label{ims5}
\end{eqnarray}
In fact, if we bear in mind that
\begin{equation}
T'=\frac{\sqrt{M'^{2}-a^{2}-Q'^{2}}}{4\pi(M'^2+M'\sqrt{M'^{2}-a^{2}-Q'^{2}}-\frac{1}{2}Q'^2)},
\end{equation}
we easily get
\begin{equation}
\frac{1}{T'}(dM'-V'_0dQ'-\Omega'_HdJ')=dS'.
\end{equation}
That is, (\ref{ims5}) is a natural result of the first law of
black hole thermodynamics.

The tunnelling rate is therefore
\begin{equation}
\Gamma\sim\exp[-2\operatorname{Im}S]=e^{\Delta S_{BH}}.
\label{spectrum}
\end{equation}
Obviously, the emission spectrum (\ref{spectrum}) deviates from
the pure thermal spectrum but consists with an underlying unitary
theory and takes the same functional form as that of uncharged
massless particles.
\section{conclusion and discussion}
With the Parikh-Wilczek tunnelling framework, We calculated the
emission rate of a charged particle tunnelling from the
Kerr-Newman black hole. The result supports Parikh-Wilczek's
conclusion, that is, the corrected spectrum is not perfect thermal
but the information is conserved during the emission process. Our
calculation also indicates that the emission process satisfies the
first law of black hole thermodynamics. In fact, the first law of
black hole thermodynamics is a combination of the energy
conservation law,
$\mathrm{d}M-V_0dQ-\Omega_H\mathrm{d}J=\mathrm{d}Q_h$ (where $Q_h$
denotes heat quantity), and the second law of thermodynamics,
$\mathrm{d}S=\frac{\mathrm{d}Q_h}{T}$. The equation of energy
conservation is suitable for any process (reversible or
irreversible process). But the equation
$\mathrm{d}S=\frac{\mathrm{d}Q_h}{T}$ is only true for a
reversible process. For an irreversible process,
$\mathrm{d}S>\frac{\mathrm{d}Q_h}{T}$. That is, Parikh-Wilczek
tunnelling framework has treated the emission process as an
reversible process. In this treatment, the black hole and the
outside approach an thermal equilibrium. There is an entropy flux
$\mathrm{d}S=\frac{\mathrm{d}Q_h}{T}$ between the black hole and
the outside. As the black hole radiate, the entropy of the black
hole decreases, but the total entropy of the black hole and the
outside is constant, and the information is conserved. Therefore,
the information conservation is a natural result of the reversible
process and the first law of black hole thermodynamics.
\acknowledgments This research is supported by National Natural
Science Foundation of China (Grant No. 10373003) and National
Basic Research Program of China (Grant No. 2003CB716300).


\begin{references}
\bibitem{Parikh1} M. K. Parikh, F. Wilczek, \textit{Phys. Rev. Lett.},
{\bf 85}, 5042(2000) [arxiv: hep-th/9907001].
\bibitem{Parikh2} M. K. Parikh, \textit{Int. J. Mod. Phys.} D {\bf 13},2355(2004) [arXiv: hep-th/0405160].
\bibitem{Parikh3} M. K. Parikh, arXiv: hep-th/0402166,
\bibitem{Hemming} S. Hemming, E. Keski-Vakkuri, \textit{Phys. Rev.} {\bf D64},
044006(2001).
\bibitem{Medved} A. J. M. Medved,  \textit{Phys.
Rev.} {\bf D66}, 124009(2002).
\bibitem{Alves} M. Alves, \textit{Int. J. Mod. Phys.} D {\bf10},
575(2001).
\bibitem{Vagenas1} E. C. Vagenas, \textit{Mod. Phys. Lett.} A{\bf 17},609(2002).
\bibitem{Vagenas2} E. C. Vagenas, \textit{Phys. Lett.} B {\bf 533},302(2002).
\bibitem{Vagenas3} E. C. Vagenas, \textit{Mod. Phys. Lett.} A{\bf
20},2449(2005).
\bibitem{Vagenas4} M. Arzano, A.J.M. Medved, E. C. Vagenas, \textit{J. High Energy Phys.}, JHEP09(2005)037.
\bibitem{Zhang1} J. Zhang, Z. Zhao, \textit{J. High Energy Phys.}, JHEP10(2005)055
\bibitem{Zhang2} J. Zhang, Z. Zhao, \textit{Phys. Lett.} B {\bf618},
14(2005).
\bibitem{Zhang3} J. Zhang, Z. Zhao, \textit{Mod. Phys. Lett.} A
{\bf20},1673(2005).
\bibitem{Zhang4} J. Zhang, Z. Zhao, \textit{Nucl. Phys.} B
{\bf725},173(2005).
\bibitem {Newman}E. T. Newman and A. I. Janis. \textit{J. Math. Phys}.
\textbf{6}, 915(1965)
\bibitem {Wald}R. M. Wald, General Relativity, Chicago and London: the
university of Chicago Press, 1984
\bibitem {Makela}J. M\"{a}kel\"{a} and P. Repo, \textit{Phys. Rev.}
\textbf{D57}, 4899(1998)
\bibitem {Damour}T. Damour, \textit{Phys. Rev.} \textbf{D18}, 18(1978)





\end{references}
\end{document}